\documentclass[aps,prl,preprint,tightenlines,superscriptaddress,showpacs,byrevtex]{revtex4}
\usepackage{graphicx}
\usepackage{dcolumn}
\usepackage{bm}

\begin{document}

\vspace*{-3\baselineskip}
\resizebox{!}{3cm}{\includegraphics{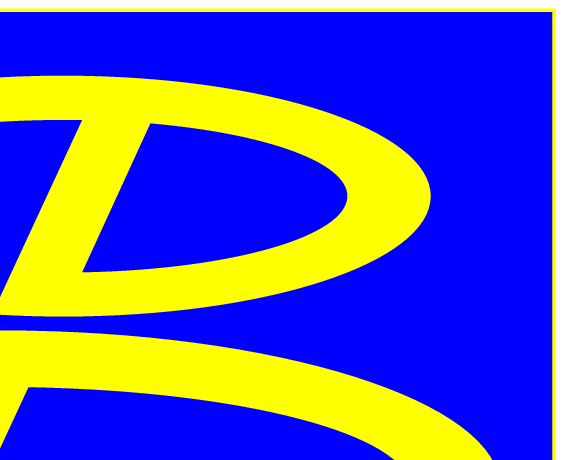}}

\preprint{KEK Preprint 2002-6}
\preprint{Belle Preprint 2002-8}

\newcommand{\svalue}{-1.21}
\newcommand{\sstaterr}{^{+0.38}_{-0.27}}
\newcommand{\ssyserr}{^{+0.16}_{-0.13}}
\newcommand{\sresult}{\svalue~\sstaterr({\rm stat})~\ssyserr({\rm syst})}
\newcommand{\cvalue}{+0.94}
\newcommand{\cstaterr}{^{+0.25}_{-0.31}}
\newcommand{\csyserr}{\pm 0.09}
\newcommand{\cresult}{\cvalue~\cstaterr({\rm stat})~\csyserr({\rm syst})}
\newcommand{\avalue}{\cvalue}
\newcommand{\astaterr}{\cstaterr}
\newcommand{\asyserr}{\csyserr}
\newcommand{\aresult}{\cresult}
\newcommand{\rt}{\rightarrow}
\newcommand{\pipi}{\pi\pi}
\newcommand{\spipi}{{\cal S}_{\pipi}}
\newcommand{\apipi}{{\cal A}_{\pipi}}
\newcommand{\skpi}{{\cal S}_{K\pi}}
\newcommand{\akpi}{{\cal A}_{K\pi}}
\newcommand{\taub}{{\tau}_{B^0}}
\newcommand{\bz}{B^0}
\newcommand{\bzb}{\overline{B}{}^0}
\newcommand{\dmd}{\Delta m_d}
\title{Study of
     {\boldmath $CP$}-Violating Asymmetries in $B^0 \to \pi^+\pi^-$ Decays}

\affiliation{Aomori University, Aomori}
\affiliation{Budker Institute of Nuclear Physics, Novosibirsk}
\affiliation{Chiba University, Chiba}
\affiliation{Chuo University, Tokyo}
\affiliation{University of Cincinnati, Cincinnati OH}
\affiliation{University of Frankfurt, Frankfurt}
\affiliation{Gyeongsang National University, Chinju}
\affiliation{University of Hawaii, Honolulu HI}
\affiliation{High Energy Accelerator Research Organization (KEK), Tsukuba}
\affiliation{Hiroshima Institute of Technology, Hiroshima}
\affiliation{Institute for Cosmic Ray Research, University of Tokyo, Tokyo}
\affiliation{Institute of High Energy Physics, Chinese Academy of Sciences, Beijing}
\affiliation{Institute of High Energy Physics, Vienna}
\affiliation{Institute for Theoretical and Experimental Physics, Moscow}
\affiliation{J. Stefan Institute, Ljubljana}
\affiliation{Kanagawa University, Yokohama}
\affiliation{Korea University, Seoul}
\affiliation{Kyoto University, Kyoto}
\affiliation{Kyungpook National University, Taegu}
\affiliation{IPHE, University of Lausanne, Lausanne}
\affiliation{University of Ljubljana, Ljubljana}
\affiliation{University of Maribor, Maribor}
\affiliation{University of Melbourne, Victoria}
\affiliation{Nagoya University, Nagoya}
\affiliation{Nara Women's University, Nara}
\affiliation{National Kaohsiung Normal University, Kaohsiung}
\affiliation{National Lien-Ho Institute of Technology, Miao Li}
\affiliation{National Taiwan University, Taipei}
\affiliation{H. Niewodniczanski Institute of Nuclear Physics, Krakow}
\affiliation{Nihon Dental College, Niigata}
\affiliation{Niigata University, Niigata}
\affiliation{Osaka City University, Osaka}
\affiliation{Osaka University, Osaka}
\affiliation{Panjab University, Chandigarh}
\affiliation{Peking University, Beijing}
\affiliation{Princeton University, Princeton NJ}
\affiliation{RIKEN BNL Research Center, Brookhaven NY}
\affiliation{Saga University, Saga}
\affiliation{Seoul National University, Seoul}
\affiliation{Sungkyunkwan University, Suwon}
\affiliation{University of Sydney, Sydney NSW}
\affiliation{Tata Institute of Fundamental Research, Bombay}
\affiliation{Toho University, Funabashi}
\affiliation{Tohoku Gakuin University, Tagajo}
\affiliation{Tohoku University, Sendai}
\affiliation{University of Tokyo, Tokyo}
\affiliation{Tokyo Institute of Technology, Tokyo}
\affiliation{Tokyo Metropolitan University, Tokyo}
\affiliation{Tokyo University of Agriculture and Technology, Tokyo}
\affiliation{Toyama National College of Maritime Technology, Toyama}
\affiliation{University of Tsukuba, Tsukuba}
\affiliation{Utkal University, Bhubaneswer}
\affiliation{Virginia Polytechnic Institute and State University, Blacksburg VA}
\affiliation{Yokkaichi University, Yokkaichi}
\affiliation{Yonsei University, Seoul}

\author{K.~Abe}               
\affiliation{High Energy Accelerator Research Organization (KEK), Tsukuba}
\author{K.~Abe}               
\affiliation{Tohoku Gakuin University, Tagajo}
\author{T.~Abe}               
\affiliation{Tohoku University, Sendai}
\author{I.~Adachi}            
\affiliation{High Energy Accelerator Research Organization (KEK), Tsukuba}
\author{Byoung~Sup~Ahn}       
\affiliation{Korea University, Seoul}
\author{H.~Aihara}            
\affiliation{University of Tokyo, Tokyo}
\author{M.~Akatsu}            
\affiliation{Nagoya University, Nagoya}
\author{Y.~Asano}             
\affiliation{University of Tsukuba, Tsukuba}
\author{T.~Aso}               
\affiliation{Toyama National College of Maritime Technology, Toyama}
\author{V.~Aulchenko}         
\affiliation{Budker Institute of Nuclear Physics, Novosibirsk}
% ...

\collaboration{The Belle Collaboration}
\noaffiliation

\date{\today}

\begin{abstract}
We present a measurement of
$CP$-violating asymmetries in $B^0 \rightarrow \pi^+\pi^-$ decays
based on
a $41.8~{\rm fb}^{-1}$ data sample collected at the $\Upsilon(4S)$ resonance
with the Belle detector at the KEKB asymmetric-energy $e^+e^-$ collider.
We fully reconstruct one neutral $B$ meson
as a  $B^0 \rightarrow \pi^+\pi^-$
$CP$ eigenstate and identify
the flavor of the accompanying $B$ meson
from its
decay products.
From the asymmetry in the
distribution of the time intervals between the two $B$ meson decay points,
we obtain the $CP$-violating asymmetry parameters
$\spipi = \sresult$
and
$\apipi = \cresult.$
\end{abstract}


\maketitle
Kobayashi and Maskawa (KM) proposed, in 1973, a model
where $CP$ violation is
incorporated as an irreducible complex phase in the
weak-interaction quark mixing matrix~\cite{KM}.
Recent measurements of the $CP$-violating parameter $\sin 2\phi_1$ 
by the Belle~\cite{CP1_Belle} and BaBar~\cite{CP1_BaBar} collaborations 
established $CP$ violation in the neutral $B$ meson system that
is consistent with KM expectations.
Measurements of other $CP$-violating parameters provide important
tests of the KM model. 
In this Letter 
we describe a measurement of $CP$-violating
asymmetries
in the mode $\bz \to \pi^+\pi^-$~\cite{CC}; 
these are sensitive to the parameter 
$\sin 2\phi_2$~\cite{alpha}.

The KM model predicts
$CP$-violating asymmetries in the time-dependent
rates for  $B^0$ and $\bzb$
decays to a common $CP$ eigenstate, $f_{CP}$~\cite{Sanda}.
In the decay chain $\Upsilon(4S)\to \bz\bzb \to f_{CP}f_{\rm tag}$,
where one of $B$ mesons decays at time $t_{CP}$ to $f_{CP}$ 
and the other decays at time $t_{\rm tag}$ to a final state
$f_{\rm tag}$ that distinguishes between $B^0$ and $\bzb$, 
the decay rate has a time dependence
given by~\cite{CPVrev}
\begin{eqnarray}
\label{eq:R_q}
{\cal P}_{\pi\pi}^q(\Delta{t}) = 
\frac{e^{-|\Delta{t}|/{\taub}}}{4{\taub}}
\left[1 + q\cdot 
\left\{ \spipi\sin(\dmd\Delta{t})   \right. \right. \nonumber \\
\left. \left.
   + \apipi\cos(\dmd\Delta{t})
\right\}
\right],
\end{eqnarray}
where $\taub$ is the $B^0$ lifetime, $\dmd$ is the mass difference 
between the two $B^0$ mass
eigenstates, $\Delta{t}$ = $t_{CP}$ $-$ $t_{\rm tag}$, and
the $b$-flavor charge $q$ = +1 ($-1$) when the tagging $B$ meson
is a $B^0$ 
($\bzb$).
The $CP$-violating parameters $\spipi$ and $\apipi$ 
defined in Eq.~(\ref{eq:R_q}) are expressed by
$\spipi = 2Im \lambda/(|\lambda|^2 + 1)$ and
$\apipi = (|\lambda|^2 - 1)/(|\lambda|^2 + 1)$,
where $\lambda$ is a complex 
parameter that depends on both $\bz\bzb$
mixing and on the amplitudes for $\bz$ and $\bzb$ decay to 
$\pi^+\pi^-$. In the SM, to a good approximation,
$|\lambda|$ is equal to the absolute value
of the ratio of the $\bzb$ to $\bz$ decay amplitudes.
%
We would have
$\spipi = \sin 2\phi_2$ and
$\apipi =0$, or equivalently $|\lambda| = 1$,
if the $b \to u$ tree amplitude were dominant.
The situation is complicated
by the possibility 
of significant contributions from gluonic $b\to d$
penguin amplitudes that
have a different weak phase and additional strong 
phases~\cite{pipipenguin}.
As a result, 
$\spipi$ may not be equal to $\sin2\phi_2$ and
direct $CP$ violation, $\apipi \neq 0$,  may occur.

This measurement 
is based on a $41.8~{\rm fb}^{-1}$ data sample,
which contains 44.8 million $B\overline{B}$ pairs, 
collected  with
the Belle detector at the KEKB asymmetric-energy
$e^+e^-$ (3.5 on 8~GeV) collider~\cite{KEKB}
operating at the $\Upsilon(4S)$ resonance.
At KEKB, the $\Upsilon(4S)$ is produced
with a Lorentz boost of $\beta\gamma=0.425$ nearly along
the electron beamline ($z$).
Since the $B^0$ and $\bzb$ mesons are approximately at 
rest in the $\Upsilon(4S)$ center-of-mass system (cms),
$\Delta t$ can be determined from the displacement in $z$ 
between the $f_{CP}$ and $f_{\rm tag}$ decay vertices:
$\Delta t \simeq (z_{CP} - z_{\rm tag})/\beta\gamma c
 \equiv \Delta z/\beta\gamma c$.

The Belle detector~\cite{Belle} is a large-solid-angle
spectrometer that
consists of a silicon vertex detector (SVD),
a central drift chamber (CDC), an array of
aerogel threshold \v{C}erenkov counters (ACC), 
time-of-flight
scintillation counters (TOF), and an electromagnetic calorimeter
comprised of CsI(Tl) crystals (ECL)  located inside 
a super-conducting solenoid coil that provides a 1.5~T
magnetic field.  An iron flux-return located outside of
the coil is instrumented to detect $K_L^0$ mesons and to identify
muons (KLM).  

\begin{figure}[h]
\begin{center}
\resizebox{0.6\textwidth}{!}{\includegraphics{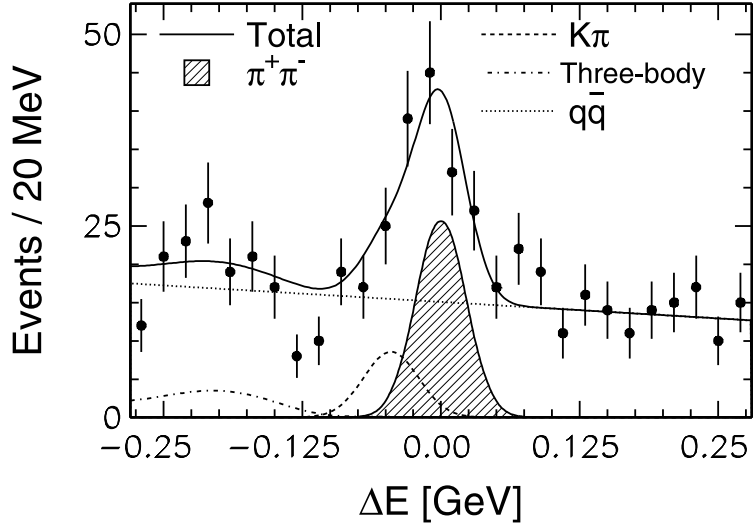}}
\end{center}
\caption{ $\Delta E$ distribution for $\pi^+\pi^-$ 
event candidates that are in the
 $M_{\rm bc}$ signal region.}
\label{fig:DeltaE}
\end{figure}

The $B^0 \to \pi^+\pi^-$ event selection is described
in detail elsewhere~\cite{pipi}.  
We use oppositely charged track pairs
that are positively identified as pions according to the
combined information from the ACC and the CDC $dE/dx$ measurement.
Candidate $B$ mesons are reconstructed using
the energy difference 
$\Delta E\equiv E_B^{\rm cms} - E_{\rm beam}^{\rm cms}$
and the beam-energy constrained
mass $M_{\rm bc}\equiv\sqrt{(E_{\rm beam}^{\rm cms})^2-(p_B^{\rm cms})^2}$,
where $E_{\rm beam}^{\rm cms}$ is the cms beam energy,
and $E_B^{\rm cms}$ and $p_B^{\rm cms}$ are the cms energy and momentum
of the $B$ candidate.
The signal region is defined as 
5.271 $< M_{\rm bc} <$ 5.287 GeV/$c^2$
and $|\Delta E |< $ 0.067 GeV, corresponding to $\pm 3\sigma$ from
the central values.
In order to suppress background from the $e^+e^- \rightarrow q\overline{q}$
continuum ($q = u,~d,~s,~c$),  we form signal and background
likelihood functions, ${\cal L}_S$ and ${\cal L}_{BG}$, 
from two variables. One is a Fisher
discriminant determined from six modified Fox-Wolfram
moments~\cite{SFW};
the other is 
the $B$ flight direction in the cms, 
with respect to the $z$ axis ($\cos\theta_B$).
We determine ${\cal L}_S$ from Monte Carlo (MC)
and ${\cal L}_{BG}$ from data, and
require ${\cal L}_S/({\cal L}_S+{\cal L}_{BG}) > 0.825$
for candidate events.
Figure~\ref{fig:DeltaE} shows the $\Delta E$ distribution for
$\pi^+\pi^-$ candidates.  
The signal yield is extracted by fitting
the $\Delta E$ distribution with a Gaussian $\pi^+\pi^-$ signal function, 
plus contributions from misidentified $B^0\rt K^+\pi^-$ events, 
three-body $B$-decays, and continuum background.
From the fit, we obtain
73.5 $\pm$ 13.8(stat) $\pi^+\pi^-$ events, 
28.4 $\pm$ 12.5(stat) $K^+\pi^-$ events, and
98.7 $\pm$ 7.0(stat) continuum events in the 
signal region. The $K^+\pi^-$ contamination is consistent
with the $K\rt\pi$ misidentification probability measured
independently. The contribution from
three-body $B$-decays is negligibly small in the signal region.

Leptons, charged pions, and kaons
that are not associated with the reconstructed
$B^0 \rightarrow \pi^+\pi^-$ decay are used to identify
the flavor of the accompanying $B$ meson.
We use two parameters, $q$ and $r$, to represent the tagging information.
The first, $q$, is already defined in Eq.~(\ref{eq:R_q}).
The parameter $r$ is an event-by-event,
MC-determined flavor-tagging dilution factor 
that ranges from $r=0$ for no flavor
discrimination to $r=1$ for {unambiguous flavor assignment.
It is used only to sort data into six $r$ intervals
(boundaries at 0.25, 0.5, 0.625, 0.75 and 0.875).
The wrong tag fractions for the six $r$ intervals, $w_l\ (l=1,6)$,
are determined 
from the data;
we use the same $w_l$ values
that were used for the $\sin 2\phi_1$ measurement~\cite{CP1_Belle}.

The vertex positions for the $\pi^+\pi^-$ and $f_{\rm tag}$ decays are
reconstructed using tracks 
with associated hits in the SVD.
Each vertex position is also constrained by
the interaction point profile smeared in the
$r$-$\phi$ plane by the average transverse $B$ meson decay length.
The $f_{\rm tag}$ vertex
is determined from all well reconstructed tracks, 
excluding the $B^0 \to \pi^+\pi^-$ candidate and
tracks that form a $K_S^0$ candidate.

The proper-time interval resolution for the signal, $R_{\rm sig}(\Delta t)$, 
is obtained by convolving a sum of
two Gaussians (a {\it main} component,
plus a {\it tail} component caused by poorly reconstructed tracks)
with a function that takes into account  
the cms motion of the $B$ mesons.
We use the same parameters
as those used for the $\sin 2\phi_1$ measurement;
the fraction and the typical width of the main Gaussian
are $0.97\pm 0.02$ and 1.49 ps, respectively~\cite{CP1_Belle}.
The background resolution function $R_{q\overline{q}}(\Delta t)$,
which is dominated by continuum background,
has the same functional
form but the parameters are obtained from a sideband region
in $M_{bc}$ and $\Delta E$.
Using these resolution functions,
we perform a $\bz$ lifetime measurement that yields
$\taub = 1.49\pm 0.21$(stat) ps for $B^0 \to \pi^+\pi^-$ candidates,
which is consistent with the world average value~\cite{PDG}.

We determine $CP$ violation parameters
by performing an
unbinned maximum-likelihood fit of a $CP$-violating
probability density function (pdf) to the 
$\Delta t$ distributions.
We define the likelihood value for each event as a
function of $\spipi$ and $\apipi$:
\begin{eqnarray}
P_i =
\int 
[\{f_{\pi\pi}^l{\cal P}_{\pi\pi}^q(\Delta t^\prime, w_l;\spipi, \apipi) +
f_{K\pi}^l{\cal P}_{K\pi}^q(\Delta t^\prime, w_l)\}
\cdot R_{\rm sig}(\Delta t_i-\Delta t^\prime) 
\nonumber \\
+ f_{q\overline{q}}^l{\cal P}_{q\overline{q}}(\Delta t^\prime)
\cdot R_{q\overline{q}}(\Delta t_i-\Delta t^\prime)]d\Delta t^\prime.
\label{eq:likelihood}
\end{eqnarray}
%
Here $f_{\pi\pi}^l$, $f_{K\pi}^l$, and 
$f_{q\overline{q}}^l$ ($= 1 - f_{\pi\pi}^l - f_{K\pi}^l$) are the fractions 
of $\pi^+\pi^-$ signal,  
$K^+\pi^-$ background, and continuum background in flavor-tagging interval
$l$, respectively.  
These fractions are determined on an event-by-event basis
as a function of $\Delta E$ and $M_{\rm bc}$, properly normalized
by the average signal and background fractions in the
signal region.
The average fractions of $q\overline{q}$ background
for six $r$ bins $(l=1,6)$ are 0.632, 0.505, 0.462, 0.440, 0.322 and 0.117.
For higher $r$ values where we are more sensitive to the
asymmetry, the fraction of continuum background decreases;
the ratio of $\pi^+\pi^-$ signal events to background
$K^+\pi^-$ events is the same for all $r$ bins.
The pdfs for $\pi^+\pi^-$ (${\cal P}_{\pi\pi}^q$), 
$K^+\pi^-$ (${\cal P}_{K\pi}^q$), and continuum background 
(${\cal P}_{q\overline{q}}$),
are convolved with their respective resolution functions. 
We use the same vertex resolution 
function for $\pi^+\pi^-$ and $K^+\pi^-$ candidates.
For the $\pi^+\pi^-$ signal,
the pdf is given by Eq.~(\ref{eq:R_q}) with $q$ replaced by
$q(1-2w_l)$, to account for the dilution due to
wrong flavor tagging.
The pdf for the $K^+\pi^-$ background is
${\cal P}_{K\pi}^q(\Delta t,w_l)
={e^{-|\Delta t|/\taub}}/{4\taub}
\{ 1 + q\cdot(1-2w_l)
  \akpi\cos(\dmd\Delta{t}) \}$,
where $\akpi$ is the $\bzb \to K^-\pi^+$ and
$B^0 \to K^+\pi^-$ decay rate asymmetry.
We fix $\akpi = 0$~\cite{pipi}, 
and $\taub$ and $\dmd$ to their world average
values~\cite{PDG}.
The pdf used for the $q\overline{q}$ background is
${\cal P}_{q\overline{q}}(\Delta t)
=\{f_\tau e^{-|\Delta t|/\tau_{\rm bkg}}/2
\tau_{\rm bkg}+(1-f_\tau)\delta(\Delta t)\}/2,$
where $f_\tau$ is the background fraction
with an effective lifetime $\tau_{\rm bkg}$ and $\delta$ is the Dirac delta
function.  We determine
$f_{\tau} = 0.011\pm 0.004$ and $\tau_{\rm bkg}= 2.7^{+1.0}_{-0.7}$ ps
from the sideband data.
In the fit,
$\spipi$ and $\apipi$ are free parameters 
determined by maximizing
the likelihood function
${\cal L}=\prod_i P_i$, where the product is over all
$B^0 \rightarrow \pi^+\pi^-$ candidates.

The result of the fit to the
162 candidates (92 $B^0$- and 70 $\bzb$-tags) that 
remain after flavor tagging and vertex reconstruction is:
\begin{eqnarray}
\spipi &=& \sresult;
 \nonumber \\
\apipi &=& \aresult.
 \nonumber
\end{eqnarray}
The result is 1.3$\sigma$ away from the physical boundary
$\spipi^2+\apipi^2 = 1$, which is consistent with
a statistical fluctuation.
The correlation between $\spipi$ and $\apipi$ is found to be 0.28.
In Figs.~\ref{fig:asym}(a) and (b), we show the
$\Delta t$ distributions for
$B^0$- and $\bzb$-tagged events together
with the fit curves;
the background-subtracted $\Delta t$ distributions
are shown in Fig.~\ref{fig:asym}(c).
Figure~\ref{fig:asym}(d) shows the background-subtracted
$CP$ asymmetry between
the $B^0$- and
$\bzb$-tagged events 
as a function of $\Delta t$. The result
of the fit is superimposed and shown by the solid curve.
\begin{figure}[!htbp]
\begin{center}
\resizebox{0.6\textwidth}{!}{\includegraphics{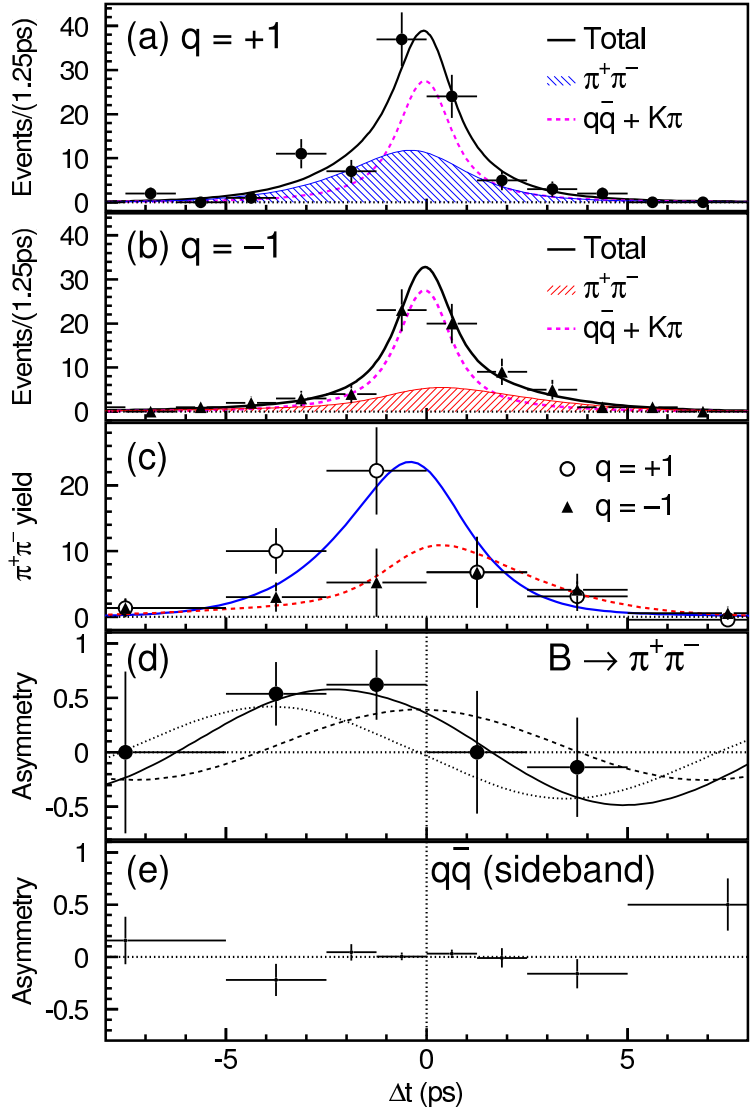}}
\end{center}
\caption{
The $\Delta t$ distributions 
 for the $B^0 \rightarrow \pi^+\pi^-$ candidates in the
signal region: 
(a) candidates with $q = +1$, i.e. the tag side is identified as $B^0$;
(b) candidates with $q = -1$; 
(c) $\pi^+ \pi^-$ yields after background subtraction. The errors
are statistical only and do not include the error of the
subtracted background obtained by a fit. The rightmost
(leftmost) bin ranges from 5 to 10 ps ($-5$ to $-10$ ps);
(d) the $CP$ asymmetry for $B^0 \rightarrow \pi^+\pi^-$
after background subtraction. The
point in the rightmost bin has a large negative value
that is outside of the range of the histogram;
(e) the raw asymmetry for $B^0 \to \pi^+\pi^-$ sideband events.
In Figs. (a) through (c), the curves show the
results of the unbinned maximum likelihood fit.
In Fig. (d), the solid curve shows the resultant $CP$ asymmetry,
while the dashed (dotted) curve is the contribution from
the cosine (sine) term.
}
\label{fig:asym}
\end{figure}

The systematic error on $\spipi$ is primarily due to
uncertainties in the background fractions ($\pm 0.09$) and
a possible fit bias near the physical boundary ($^{+0.11}_{-0.02}$). 
For $\apipi$, the background fractions ($\pm 0.06$) and the wrong-tag
fractions ($\pm 0.06$) are the two leading components.
We find that uncertainties in 
$f_{\pi\pi}^l$, $f_{K\pi}^l$ and $f_{q\overline{q}}^l$
in Eq.~(\ref{eq:likelihood})
account for the largest systematic errors.
We add their contributions in quadrature to obtain the above values,
where each contribution is obtained by
varying a parameter by its error and repeating the fit.
Other sources of systematic error are
uncertainties in the resolution function, physics parameters
($\dmd$, $\taub$, and $\akpi$) and the background modeling.
A value of $\akpi = -0.06\pm 0.08$ is obtained from the self-tagged 
$B^0\to K^+ \pi^-$   
sample~\cite{pipi}; this introduces a systematic error of
$<0.01$ for $\spipi$ and
$^{+0.02}_{-0.01}$ for $\apipi$.

We validate our fitting procedure using a large ensemble of MC
pseudo-experiments wherein events are generated with nominal pdfs and the
observed number of events. For various input values of
$\spipi$ and $\apipi$, we confirm that there is no bias in
the fit results.
We find that the average expected errors,
0.52 for $\spipi$ and 0.35 for $\apipi$,
are larger than our measurements.
However, the probability of obtaining errors that are smaller
than our measurement is 5.4\% for $\spipi$ and
23.6\% for $\apipi$; the results are within 
the expected range of statistical
fluctuations. 
The measured likelihood is
in good agreement with the average likelihood value
obtained in the pseudo-experiments.

We perform a number of cross checks. 
We examine the event yields and $\Delta t$ distributions for
$B^0$- and $\bzb$-tagged events in the sideband region and find
no significant asymmetry as shown in Fig.~\ref{fig:asym}(e).
We select $B^0 \to K^+\pi^-$ candidates,
which have the same track topology as $B^0 \to \pi^+\pi^-$,
by positively identifying charged kaons.
A fit to 309 candidates ($209 \pm 16$ signal events)
yields 
$\taub = 1.73\pm 0.15$(stat) ps and $\dmd = 0.57 \pm 0.08$(stat) ps$^{-1}$;
these are consistent with the world average values~\cite{PDG}.
$\akpi$ is $0.07 \pm 0.17$, in agreement with the
counting analysis mentioned above
and $\skpi = 0.15 \pm 0.24$, which is consistent with zero.
We also select $B^0 \to D^-\pi^+$, $D^{*-}\pi^+$ and $D^{\*-}\rho^+$
candidates using the same event shape criteria.
Neither mixing-induced nor direct $CP$-violating asymmetry is observed
as expected.
As an additional test of the consistency of the background treatment,
we add events from the $B^0 \to \pi^+\pi^-$ sideband and
adjust their $\Delta E$ and $M_{\rm bc}$ values.
A fit to this background-enriched control sample,
which has a background fraction comparable to the $B^0 \to \pi^+\pi^-$ sample,
yields
${\cal S} = 0.08 \pm 0.06$ and
${\cal A} = 0.03 \pm 0.04$,
both consistent with a null asymmetry.

We determine the statistical significance from the likelihood
function, taking into account the boundary of the physical region
as well as the effect of the systematic error.
The Feldman-Cousins frequentist approach~\cite{FeldmanCousins}
gives a 99.6\% confidence level (C.L.) for 
$-1 \leq \spipi < 0$, equivalent to 
a $2.9\sigma$ significance for a Gaussian error.
A similar analysis yields a significance of $2.9\sigma$ for 
$0 < \apipi \leq 1$.
The 95\% C.L. intervals are found to be 
$-1.00 \leq \spipi < -0.39$
and
$+0.30 < \apipi \leq +1.00$,
respectively~\cite{ToyMC}. 

In summary, we have measured the
$CP$ violation parameters in $B^0 \rightarrow \pi^+\pi^-$ decay.
Our result for $\spipi$ indicates that mixing-induced $CP$
violation
is large. The large $\apipi$ term is an indication of direct
$CP$ violation in $B$ meson decay, and suggests that there is
a large hadronic phase and interference between the tree and
penguin amplitudes~\cite{apipi_BaBar}. 
In this case the precise determination of $\sin 2\phi_2$ from $\spipi$
requires additional measurements
including the branching fractions for the decays
$B^0 \rightarrow \pi^0 \pi^0$~\cite{GL}. 

We wish to thank the KEKB accelerator group for the excellent
operation of the KEKB accelerator.
We acknowledge support from the Ministry of Education,
Culture, Sports, Science, and Technology of Japan
and the Japan Society for the Promotion of Science;
the Australian Research Council
and the Australian Department of Industry, Science and Resources;
the National Science Foundation of China under contract No.~10175071;
the Department of Science and Technology of India;
the BK21 program of the Ministry of Education of Korea
and the CHEP SRC program of the Korea Science and Engineering Foundation;
the Polish State Committee for Scientific Research
under contract No.~2P03B 17017;
the Ministry of Science and Technology of the Russian Federation;
the Ministry of Education, Science and Sport of the Republic of Slovenia;
the National Science Council and the Ministry of Education of Taiwan;
and the U.S.\ Department of Energy.


\begin{thebibliography}{99}
\bibitem{CC}
Throughout this Letter, 
the inclusion of the charge conjugate mode decay is implied
unless otherwise stated.
\bibitem{alpha} Another naming convention, 
$\beta$ ($= \phi_1$) and $\alpha$ ($= \phi_2$), is
also used in the literature.
\bibitem{ToyMC}
An ensemble of MC pseudo-experiments indicates
a 1.6\% probability to measure
$CP$ violation at or above the one we observe
when the input values are $\spipi = \apipi =0$.
\end{thebibliography}
\end{document}